\documentstyle[epsfig,times,12pt]{article}
\def\be{\begin{equation}}
\def\ee{\end{equation}}
\def\br{\begin{eqnarray}}
\def\er{\end{eqnarray}}
\def\no{\nonumber}
\begin{document}
\title{Solar Magnetic Field Profile: A Natural Consequence of RSFP Scenario}
\author{Bhag C. Chauhan{\footnote{chauhan@cfif.ist.utl.pt}}\\
\it Centro de F\'{i}sica das Interac\c{c}\~{o}es Fundamentais (CFIF)\\
Departmento de Fisica, Instituto Superior T\'{e}cnico\\
Av. Rovisco Pais, 1049-001, Lisboa-PORTUGAL}
\date{\today}
\maketitle
\begin{abstract}
Assuming the solar neutrino deficit is resolved by the resonant interaction of the neutrino magnetic moment with the solar magnetic field --in the framework of Resonant Spin Flavour Precession (RSFP) scenario-- the solar magnetic field profile function has been derived from the scenario in the light of solar neutrino data. An approximate qualitative analysis has been done for vanishing vacuum mixing and it has been found that the profile derived is quite stable in nature. As because on changing the neutrino parameters ($\mu_{\nu},~\Delta{m^2}$) and the solar neutrino data the profile is just scaled along the axes. In principle, the nature of the profile is strongly dependent on the solar matter density distribution function. The current approach is quite different from the usual one- in which the best field profile is discovered by performing $\chi^{2}_{min.}$ calculations using solar neutrino data. Furthermore, the profile derived in the present work --when tested by $\chi^{2}_{min.}$ calculations-- was found to be the best suited one, for the solar interior.  
\end{abstract}
\newpage
\section{Introduction} 
Solar neutrino deficit has been strongly established in all the first generation solar neutrino experiments \cite{r1} and in the --already started-- second generation experiments \cite{r2}. The observed solar neutrino flux is considerably lesser than the standard solar model (SSM) predictions \cite{r3}. This leads one to assume the non-standard neutrino properties. The recognised solutions to this Solar Neutrino Problem (SNP) --based on the nonstandard neutrino properties-- are the the following: neutrino oscillations (Low MSW, VO-QVO and LMA MSW); neutrino magnetic moment solutions (resonant spin-flavour precession (RSFP) and non-resonant spin-flavour precession (NRSFP)) and the nonstandard neutrino interactions (NRIs). However the most pouplar solution to the SNP is the oscillation solution (LMA MSW), yet the recent (post-SNO) data equally favours the neutrino magnetic moment solutions \cite{r3a}. In the present work it is assumed that the RSFP \cite{r4} of solar neutrinos is solely responsible for the neutrino deficit. This scenario is based on the presence of non-zero neutrino transition magnetic moment interaction with the transverse solar magnetic field along the path of neutrino trajectory. If neutrinos are Majorana particles the RSFP converts $\nu_{eL}$ into $\overline\nu_{\mu R}$ or $\overline\nu_{\tau R}$, which are sterile for the chlorine detector but do contribute in Kamiokande/Superkamiokande and in SNO event rates through comparatively smaller neutral current cross-sections. RSFP of neutrinos can explain not only the solar neutrino deficit but also the apparent time variations (anticorrelation) \cite{r5} --not confirmed or disproved yet-- of the solar neutrino flux with the solar magnetic activity. 
\par 
Despite the unestablished argument of anticorrelation, there are several reasons which motivate RSFP and its conequences for solar neutrinos. In fact, different degrees of suppression \cite{r5a} in the survival probabilities of the low energy (pp-neutrinos), intermediate energy ($^{7}Be$,CNO,pep-neutrinos) and high energy part of $^{8}B$-neutrino sector, is the inbuilt feature of RSFP. 
Furthermore, it has been found that the event rate fits for RSFP from the solar neutrino experiments are remarkably better than those for the best oscillation solution (LMA MSW) \cite{r7} whereas the fits for recoil energy spectrum in Superkamiokande are found nearly of the same quality as that of LMA MSW \cite{r6}. By analysing the solar neutrino data within the framework of RSFP scenario it has been noted that the quality of data fit is very sensitive to the magnetic field profiles used \cite{r6}. The strongest field intensity corresponds to the lowest survival probability and vice-versa. And in this way RSFP explains the general shape of probability, which naturally appears as a reflection of the field profile. Conversely, a particular shape of the magnetic field profile can be obtained from the scenario, as a reflection of solar neutrino data. The procedure --in the present work-- of extracting the profile from the RSFP framework is more detailed and transparent than that has been done with the other authers \cite{r18}. 
\par 
In this paper, a rough estimate of the solar neutrino data \cite{r5a} has been used for a qualitative analysis and thereby the field profile function is derived. And the effects on the shape of the profile on changing the neutrino parameters and the solar neutrino data have also been investigated. The paper has been divided in four sections. The solar magnetic field, at the present status, is briefed in section-2. In section-3 the derivation of the field profile from the RSFP scenario is achieved. And, the section-4 is devoted to the results and the discussion.  

\section{Magnetic Field in the Sun}
Very little is known about the magnetic field inside the sun. Apart from the anormous success of the SSM in predicting the thermal and nuclear evolution inside the sun it does not through enough light on the solar magnetic properties.
In the most realistic MHD model of regular magnetic fields in the convective zone, the main component of field is toroidal with opposite polarities in the northern and the southern solar hemispheres. The MHD models \cite{r10} do not exclude the presence of a significant magnetic field of few hundred kilogauss at the bottom of the convective zone (CZ). The convective field is somehow caused by the relative rotation of solar layers in the convective zone. At the sunspot maximum the surface field may reach $10^{3}-10^{4}$Gauss inside the spot and in the sunspot minimum the field falls below $10^{2}$ G. Below the surface to the bottom of the CZ the field cannot be measured directly except for an upper limit of 300 kG from helioseismological data \cite{r11}. There are recent models \cite{r12} which argue that the field generation of about $100$ kG occurs in the shear layer near the bottom of CZ. A strong field of $4\times10^{5}$G extending over $3\times10^{4}$km at the bottom of the convective zone would force the CZ to extend deep enough to sufficiently distroy the $^{7}Li$ abundance though not completely destroying it during the $4.5\times10^{9}$ yrs \cite{r13}. 
\par
Below the bottom of the convective zone, the field could be large but the field response time due to plasma effect is of the order of $10^{10}$ yrs in this region \cite{r14}. Such a field, if it exists, will remain frozen over the intire life history of sun. 
Parker, in reference \cite{r10} has shown that a field in excess of $0.5\times10^{8}$G in the central core would be lost from the sun during its evolution as a consequence of its buoyancy. 
Unlike the CZ, the radiative zone (RZ) is not continuously mixed and rotates as a solid body. The properties of the magnetic fields in the interior of the sun has been investigated by some authers \cite{r16} by considering the poloidal nature of magnetic fields. According to a recent study by Friedland {\it et al.}, a toroidal field of complex spatial structure can exist in the RZ \cite{r16a}. 
\par
In the absence of reliable knowledge of solar magnetic fields from the astrophysical observations it becomes worthwhile to extract the solar magnetic field profile from the solar neutrino data within the framework of RSFP scenario. However, the bounds discussed above can offer little help to constrain the magnitude of field in the solar interior.  

\section{Derivation of Solar Magnetic Field Profile} 
In the RSFP scenario --for Majorana neutrino flavours in the chiral basis ($\nu_{e},~\nu_{\mu},~\overline\nu_{e},~\overline\nu_{\mu}$) --the neutrino propagation in the solar medium with a transverse component of magnetic field ``$B_{\perp}$'', can be described by a Schrodinger-like evolution equation  
\br
i\frac{d}{dx}\left( \begin{array}{c} \nu_{e} \\ \nu_{\mu} \\
\overline\nu_{e} \\ \overline\nu_{\mu} \end{array} \right) ~~
=~~H~ \left( \begin{array}{c} \nu_{e} \\ \nu_{\mu} \\
\overline\nu_{e} \\ \overline\nu_{\mu} \end{array} \right) ~~
\er
with the Hamiltonian matrix given by
\begin{small}
\br
H~=~\left(\begin{array}{cccc}
\frac {\Delta m^{2}}{2E} \sin ^{2}\theta + a_{\nu_{e}} & 
\frac {\Delta m^{2}}{4E}\sin 2\theta &   0   & \mu^{*} B \\
\frac {\Delta m^{2}}{4E}\sin 2\theta & 
\frac {\Delta m^{2}}{2E}\cos ^{2}\theta + a_{\nu_{\mu}} & 
-\mu^{*} B &   0  \\
0   & \mu B  & \frac {\Delta m^{2}}{2E}\sin ^{2}\theta - a_{\nu_{e}} &
\frac {\Delta m^{2}}{4E}\sin 2\theta  \\ 
-\mu B  &  0  &
\frac {\Delta m^{2}}{4E}\sin 2\theta & 
\frac {\Delta m^{2}}{2E} \cos^{2}\theta - a_{\nu_{\mu}}  \end{array} \right),  \no  
\er    
\end{small}

where $\Delta{m^2}=m_{2}^2-m_{1}^2$, $\theta$ is the vacuum mixing angle and $\mu~(=\mu_{e\mu})$ denotes the transition (off-diagonal) magnetic
moment. The diagonal magnetic moments are absent for the Majorana neutrinos as a consequence of CPT invariance. The matter potentials for a neutral unpolarised medium are
given by
\br
a_{\nu _{e}}~=~\sqrt{2}G_{F}~(N_{e}-\frac {N_{n}}{2}), ~~ \nonumber \\
a_{\nu_{\mu}}~=~-\frac {1}{\sqrt{2}}G_{F}~N_{n}~~.     \nonumber
\er
 where $G_{F}$ is the Fermi constant and $N_{e},~N_{n}$ are the number densities of electrons, neutrons, respectively, in the solar medium.     \\
\par
In the framework, the neutrino propagation is considered to be adiabatic except for a small region in the vicinity of resonance, where the two eigenvalues are the closest, there is a chance of nonadiabatic behaviour that the neutrino may jump from one eigenvalue to the other. This nonadiabatic effect is measured by the crossing probability between the eigenvalues.
For `$\theta$' vanishingly small, the spin-flavour conversion
($\nu_{e}\rightarrow\overline\nu_{\mu})$ is governed by the ``$2\times 2$''
submatrix
\br
\left(\begin{array}{cc} a_{\nu_{e}} &  \mu^{*} B  \\  -\mu B   & \frac {\Delta m^{2}}{2E}-a_{\nu_{\mu}} \end{array} \right) \nonumber
\er
At resonance, the diagonal elements of the matrix are equal
\begin{small}
\br
\frac {G_{F}}{\sqrt{2}}\frac{11}{6}~N_{e}~=~ 
 \frac {\Delta m^{2}}{2E}~+~\frac{G_{F}}{\sqrt{2}}\frac{N_{e}}{6}   
\er    
\end{small}
here $a_{\nu _{e}}~=~\frac {G_{F}}{\sqrt{2}}\frac{11}{6}~N_{e}$ and 
$a_{\nu_{\mu}}~=~-\frac{G_{F}}{\sqrt{2}}\frac{N_{e}}{6}$ with the approximation $N_{n}~\approx~N_{e}/6$. 
The equation can be further simplified as the following
\br
\frac {\Delta m^{2}}{2E}~=~\frac{5{\sqrt{2}}G_{F}N_{e}}{6}
\er
The electron density in the sun follows an exponential behaviour along the neutrino trajectory and is well approximated by \cite{r20}
$$G_{F}N_{e}~~=~~2.151 \times10^{-11} exp~(-r/0.09 R_{s})~eV $$
where $r$ is the radial distance from the sun and $R_{s}$ is the solar radius. Now, by eliminating the factor $~G_{F}N_{e}$ in eqn. (3), one can obtain 
\be
 \frac {\Delta m^{2}}{2E}~=~ 2.535 \times10^{-11} exp~(-r/0.09 R_{s})~eV 
\ee    
From equation (4), the resonance points in the solar interior can be derived as
\be
r~=~0.09~R_{s}~ln~\left( \frac {E}{\Delta m^{2}}
\times ~5.07~\times~10^{-11}~eV \right)
\ee
  
 Assuming a linear density fall off along the nonadiabatic region, the crossing probability in the Landau-Zener approximation is given by
\be
P_{LZ}~=~exp~\left( -~\pi \gamma_{RSFP} \right)
\ee
Here, $ \gamma_{RSFP}$ is the adiabaticity parameter
\be
\gamma_{RSFP}~=~\frac {2\mu^{2}[B(x)]^{2}}{\left(
\frac {\Delta m^{2}}{2E}\right)}~0.09R_{s})~, \no
\ee
 which is very sensitive to the nature of magnetic field profile $B(x)$. \\
Here $~x=r/R_{s}$, the position of the resonance points in the units of solar radius.
\par
An approximate survival probability for neutrinos, when they all are assumed to be produced at the same point, can be obtained by using the analysis of Parke \cite{r21} as 
\be
P_{sur}~=~ \frac {1}{2}~+~\left( \frac {1}{2}~-~P_{LZ}
\right)\cos{2\tilde\theta_{i}}\cos{2\tilde\theta_{f}}
\ee

where $\tilde\theta_{i}$ and $\tilde\theta_{f}$ are the RSFP mixing angles at the beginning and at the end of the neutrino trajectory and are defined as
$$\tan{2\tilde\theta}~=~\frac {4E \mu B}{\Delta m^{2}-2E a_{\nu _{e}}(0)}$$
here $a_{\nu _{e}}(0)$ is the matter potential at the production point of neutrinos. When neutrinos are born far from the resonance and then pass through it, the mixing angles can be well approximated as
$$\cos{2\tilde\theta _{i}} \cos{2\tilde\theta _{f}} ~\sim~-1$$ 
and the total survival probability is given by
\be
P_{sur}~=~exp~\left( -~\pi \frac {2\mu^{2}[B(x)]^{2}}{\left(
\frac {\Delta m^{2}}{2E}\right)}~0.09R_{s} \right)~~.
\ee
Now, using equation (4), one can eliminate the factor $\frac {\Delta m^{2}}{2E}$ from the above equation as
\be
P_{sur}~=~exp~\left( -~\pi \frac {2\mu^{2}[B(x)]^{2}}{2.535 \times10^{-11} exp~(-r/0.09 R_{s})}~0.09R_{s} \right)~~.
\ee
And in this way, the desired field profile can be extracted from eqn. (10), as
\be
B(x)~={\sqrt{\left(\frac {5.07\times10^{-11}}{4\pi \mu^{2}~0.09R_{s}}\right)}}
{\sqrt{\left(-ln(P_{sur})\right)}}~exp~(-x/0.18)~~.
\ee
the profile can be further generalised as 
\be
B(x)~=\left[B_{\mu}\right]\left[B_{P}\right]~exp~(-x/0.18)~~.
\ee
where $B_{\mu}={\sqrt{\left(\frac {5.07\times10^{-11}}{4\pi \mu^{2}~0.09R_{s}}\right)}}$ and $B_{P}={\sqrt{\left(-ln(P_{sur})\right)}}$. 

\section{Results and Discussion}
The derived profile (12) is exponential in nature which shows a strong dependence on the exponential behaviour of the solar matter density distribution used in the derivation \cite{r20}. The amplitude of the profile is a function of magnetic moment and the survival probability of neutrinos. However, the exponent of the profile is a function of ``$\Delta{m^2}$'' and neutrino energy, `` E''. The effects of the neutrino parameters and the solar neutrino data on the shape of the profile has a detail study in this section.  
For a model calculation, the solar neutrino spectrum $(\le~15~MeV)$ has been categorised in three energy bands: low energy $(\le~0.42~MeV)$; intermediate energy $(0.43~-~0.861~MeV)$ and high energy $(0.862~-~15~MeV)$. Assuming $P_{L};~P_{I};~P_{H}$ are the three central values of the survival probabilities corresponding to the three categories of the neutrinos, respectively. Now the field profile equation (12) can be rewritten as
\be
B_{1}(x_{L})~=\left[B_{\mu}\right]\left[B_{P_{L}}\right]~exp~(-x_{L}/0.18)
\ee
\be
B_{2}(x_{I})~=\left[B_{\mu}\right]\left[B_{P_{I}}\right]~exp~(-x_{I}/0.18)
\ee
\be
B_{3}(x_{H})~=\left[B_{\mu}\right]\left[B_{P_{H}}\right]~exp~(-x_{H}/0.18)
\ee

Assuming $\Delta{m^2}=1\times10^{-8}~eV^{2};~\mu_{\nu}$=$10^{-11}~\mu_{B}$ ($\mu_{B}$ is Bohr magneton) and $P_{L}=0.9;~P_{I}=0.2;~P_{H}=0.5$ [The pattern of survival probabilities taken has only a qualitative importance. However, this pattern is not supported by the recent global analysis ($P_{L}>~P_{I}>~P_{H}$)]. The field distribution has been obtained for eqns. (13-15) and shown in figure 1. 
\begin{figure}
\centering
\leavevmode\epsfysize=7cm \epsfbox{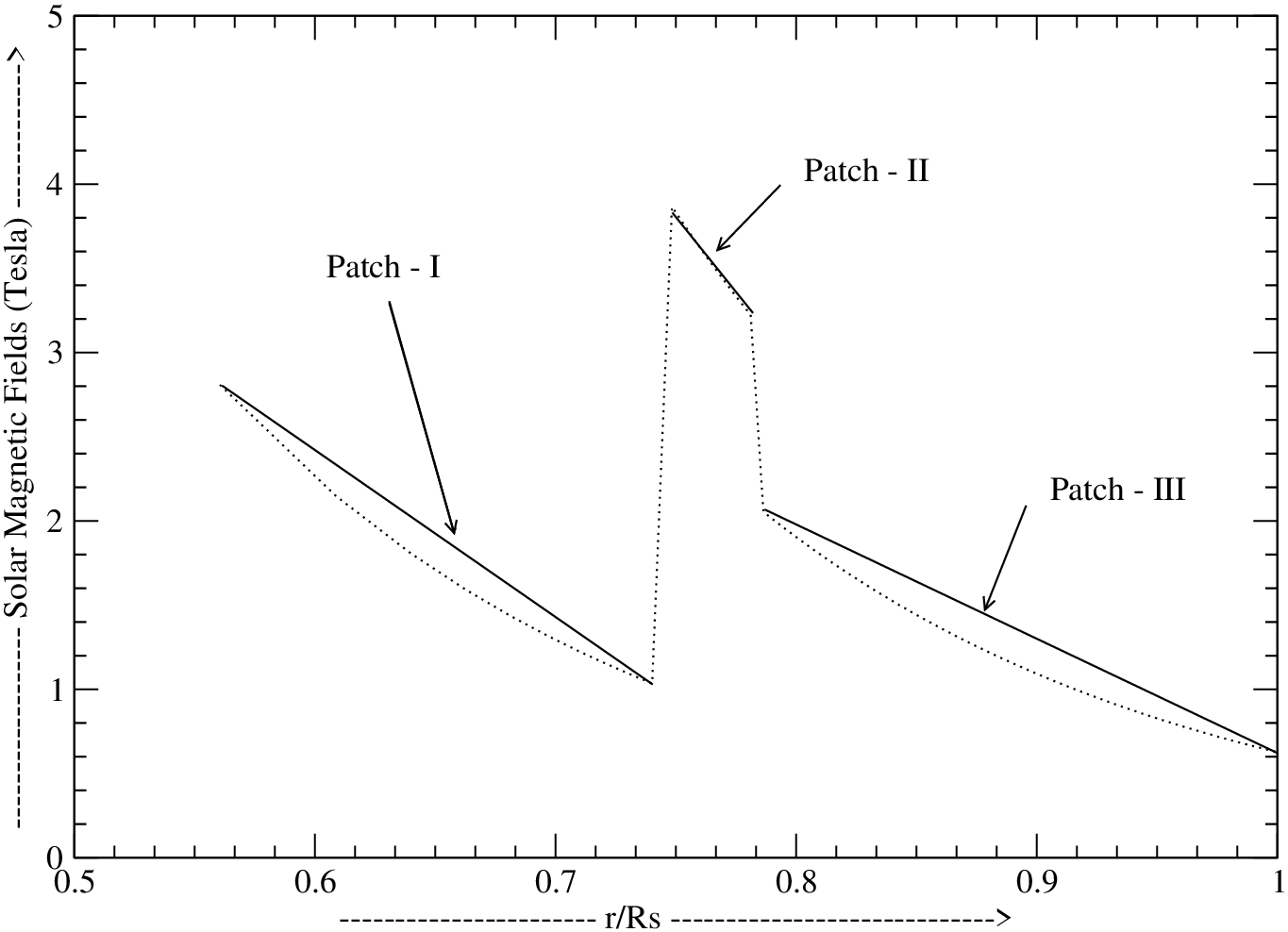}
\caption{A model solar field-patch reflection of three energy bands in the solar neutrino spectrum}
\label{fig1}
\end{figure}
The three patches of field --seen in the figure-- are as a result of one to one correspondence with the three energy bands --assumed-- in the solar neutrino spectrum.  
The first patch [$B_{1}(x_{L})$] corresponds to the resonance points of the mildly suppressed low energy neutrinos; the strongly suppressed intermediate energy neutrinos reflect the magnetic field points of the second patch [$B_{2}(x_{I})$] and the resonance points of the moderately supressed high energy neutrinos provide the field points corresponding to the third patch [$B_{3}(x_{H})$]. The strongest field corresponds to the second (middle) patch is a result of large conversion of the intermediate energy neutrinos. And the moderate field in the third patch is reflected by the moderate conversion for the high energy sector of the $^{8}B$ neutrinos. However, unlike this, less conversion of low energy neutrinos couldn't reflect a very weak magnetic field in the first patch because according to the eqn. (5) the low energy neutrinos resonate in the inner part of sun --region of higher matter density. So, larger field is required to compete with the higher effective resonance density in order to obtain small adiabatic conversion (10\%) for pp-neutrinos. 
\par
The change in the value of neutrino parameters ($\Delta{m^2},~\mu_{\nu}$) proportionally scales the field amplitude in the solar interior.    
\begin{figure}
\centering
\leavevmode\epsfysize=7cm \epsfbox{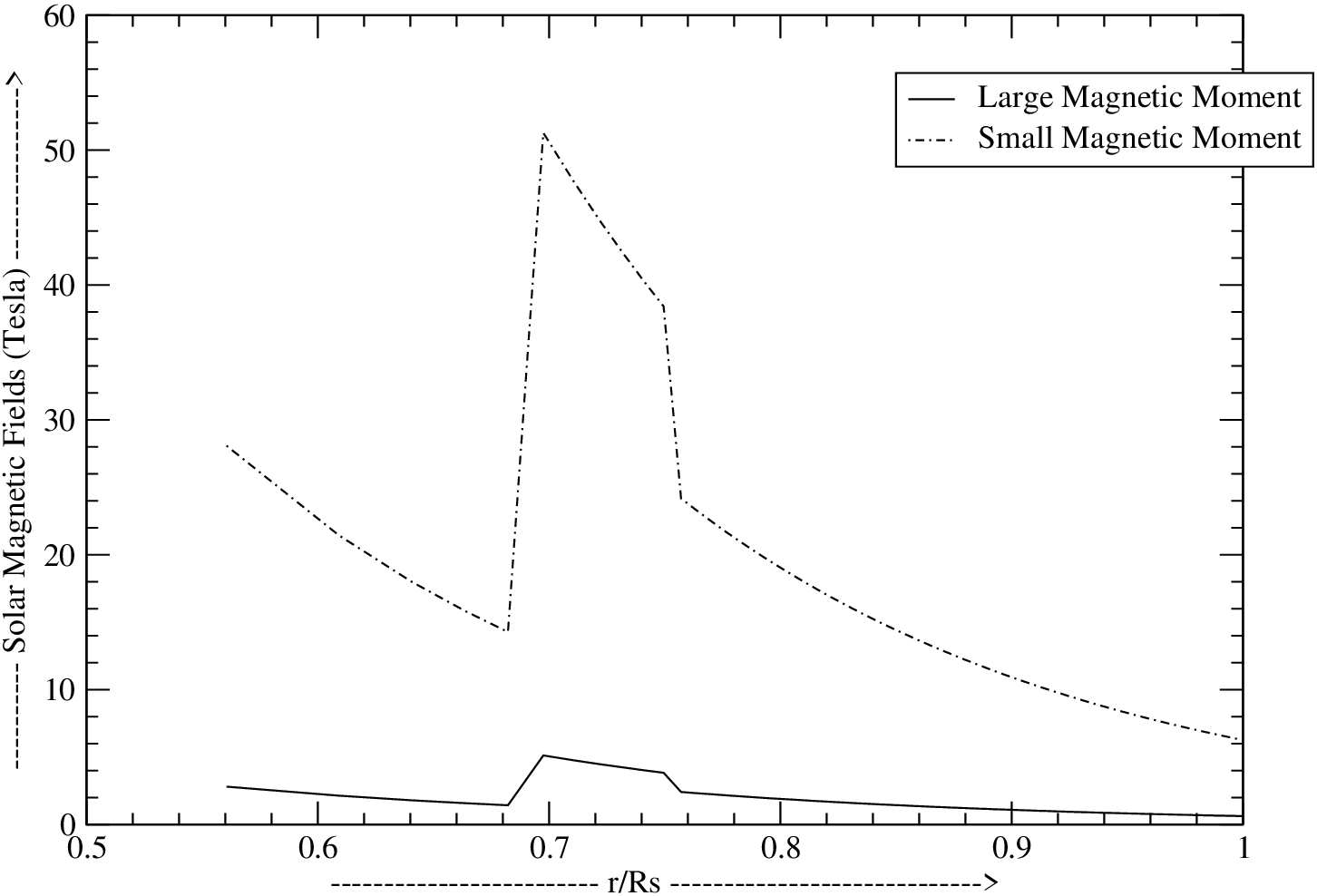}
\caption{The scaling of the field by changing the neutrino magnetic moment, $\mu_{\nu}$}
\label{fig2}
\end{figure}
The effect of the neutrino magnetic moment, for the same set of data and constant $\Delta{m^2}$, has been shown in fig. 2. It affects equally all the three patches of the field. For small magnetic moment ($\mu_{\nu}=10^{-12}~\mu_{B}$) the field is increased by one order at all the points of the profile as compared to the field distribution for large magnetic moment ($\mu_{\nu}$=$10^{-11}~\mu_{B}$). So any change in the ``$\mu_{\nu}$'' scales the magnetic field strength at all the points by the same amount. 
\begin{figure}
\centering
\leavevmode\epsfysize=7cm \epsfbox{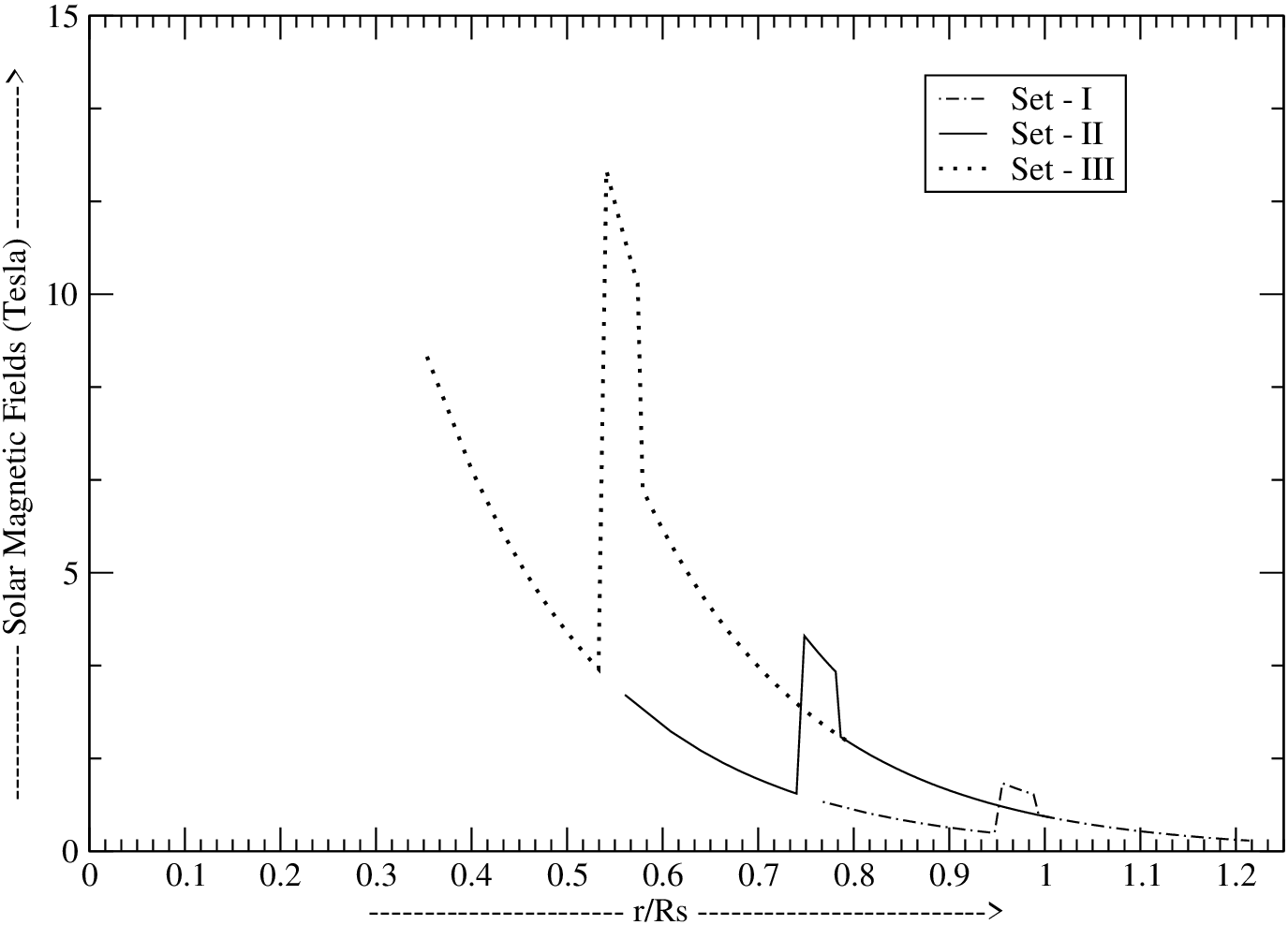}
\caption{The shifting of the field by varying $\Delta{m^2}$}
\label{fig3}
\end{figure}
\par
The change in the value of $\Delta{m^2}$ shifts the positions of resonance points of neutrinos --as can be seen from eqn. (5). The parameter ($\Delta{m^2}$) appears in eqn. (12) in such --an indirect way-- that it can only shift the field profile along the x-axis. The effect of its variation --for constant $\mu_{\nu}$ and the same set of data-- on the field profile has been shown in the figure 3.
In the figure there are shown three sets of the profile, which corresponds to the three different values of $\Delta{m^2}$. Set-I: $\Delta{m^2}=10^{-9}~eV^{2}$; Set-II: $\Delta{m^2}=10^{-8}~eV^{2}$; Set-III: $\Delta{m^2}=10^{-7}~eV^{2}$. 
For large value of $\Delta{m^2}$ as the resonance points are shifted towards the solar interior thereby the field profile is accordingly shifted to the region. For a smaller value of $\Delta{m^2}$ the profile is moved towards the surface and for $\Delta{m^2}~\le~10^{-9}~eV^{2}$ it extends beyond the solar surface as shown in the figure.  
\par
The amplitude strength of the magnetic field profile (12) is also sensitive to the solar neutrino data. The effect --for constant $\mu_{\nu}$ and $\Delta{m^2}$-- has been shown in the fig. 4. Here two arbitrary sets of data are used \\
Set-I: $P_{L}=0.9,~P_{I}=0.2,~P_{H}=0.4$  \\
Set-II: $P_{L}=0.8,~P_{I}=0.001,~P_{H}=0.6~$ \\
It has been noted that the change of survival probability of any of the neutrino category, the field is scaled as a square root of negative of logarithmic value of the probability at the same point of the corresponding field patch. So for any variation of the solar neutrino data, only the amplitude of the profile gets affected.  
\begin{figure}
\centering
\leavevmode\epsfysize=7cm \epsfbox{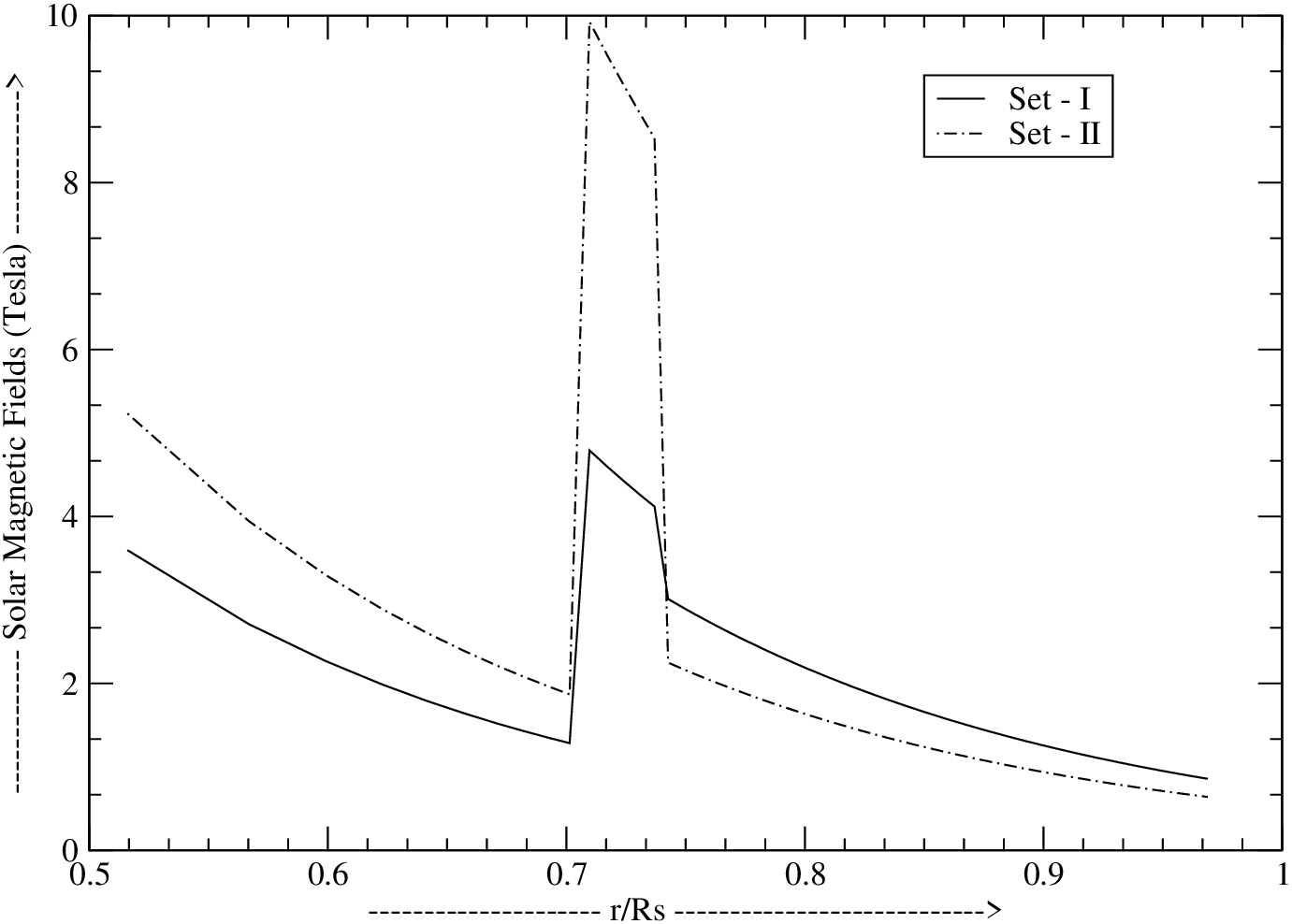}
\caption{Amplitude scaling of the field by the solar neutrino data}
\label{fig4}
\end{figure}
\par
In the present work, it has been noted that derived profile function is a hidden characteristic of RSFP scenario and it is the nature of function (e.g. exponential in the present case) which does't depend on the solar neutrino observations and the neutrino parameters. The solar neutrino data can only constrain the amplitude of the profile and the neutrino parameters can only shift the profile or scale the field amplitude along the axes. The derived profile function is found to be strongly depending on the matter density distribution of the solar interior. Furthermore, the profile can exist --radially of course-- in any part of solar interior. It is the value of $\Delta{m^2}$ which confirms its place inside the sun. For  $\Delta{m^2}~\approx~10^{-8}~eV^{2}$ the profile exists in the convective zone and for $\Delta{m^2}~\approx~10^{-5}~eV^{2}$ the profile shifts to the radiative zone of sun. The relevance of this model work remains viable whether KamLAND \cite{r22} proves LMA MSW as a solution for SNP or not. As, if LMA is confirmed as a solution, even then RSFP can survive as a sub-leading effect. And if LMA is ruled out then the chances of RSFP to exist as a dominant solution can not be ignored. It is the next awaited experiment BOREXINO  which will decide the fate of RSFP \cite{r23, r3a}. 
\newpage
\par     
To conclude, the solution to SNP based on the neutrino spin flavour flip in the transverse solar magnetic field (RSFP) is appealing from the point of view of the data fits. RSFP can happen in the relic frozen/ constant field of radiative zone (see ref.\cite{r16a} and B.C. Chauhan {\it et al.} in \cite{r3a}); in this way the generic consequences of RSFP in the convective zone of sun ( 11 years and semiannual variations of neutrino signal --which have not been observed) can be avoided. If RSFP is a true solution of Solar Neutrino Problem (SNP) it becomes more plausible to study the profile derived in the present work --from the inbuilt structure of the framework-- rather to impose a profile from outside for the $\chi^{2}$ testing. One of such excercises --of $\chi^{2}$ testing-- has been proved successful (see our paper in \cite{r3a}), recently, where we have assumed this profile (12) existing in the radiation zone of sun. And, interestingly, this profile has been found as the best suited one --in the light of the most recent (post SNO) data. 
\newpage
{\it {\bf Acknowledgements:}} \\
The author is grateful to E. Kh. Akhmedov for enlightening discussions and useful comments. The work was supported by Funda\c{c}\~{a}o para a Ci\^{e}ncia e a Tecnologia through the grant SFRH/BPD/5719/2001.

\end{document}